\documentclass[prl,reprint,amsmath,amssymb]{revtex4-1}
\usepackage[pdftex]{graphicx}
\usepackage{subfigure}
\usepackage{dcolumn}
\usepackage{epstopdf}
\usepackage{color}
\usepackage{upgreek}
\usepackage[hidelinks]{hyperref}
\usepackage{bm}
\makeatletter
\def\btt#1{\texttt{\@backslashchar#1}}%
\DeclareRobustCommand\bblash{\btt{\@backslashchar}}%
\makeatother
\newcommand{\SD}[1]{\textcolor{black}{#1}}
\newcommand{\SK}[1]{\textcolor{black}{#1}}

\newcommand{\SN}[1]{\textcolor{black}{#1}}
\begin{document}

\title{Giant electroviscous effects in a ferroelectric nematic liquid crystal}
\date{\today}
\author{M. Praveen Kumar$^{1}$, Jakub Karcz$^{2}$, Przemyslaw
 Kula$^{2}$,   Smarajit Karmakar$^{3}$ and Surajit Dhara$^{1}$}
\email{surajit@uohyd.ac.in} 
\affiliation{$^1$School of Physics, University of Hyderabad, Hyderabad-500046, India\\
$^{2}$Institute of Chemistry, Faculty of Advanced Technologies and Chemistry, Military University of Technology, Warsaw, Poland.\\
 $^{3}$Tata Institute of Fundamental Research, Hyderabad, 500107, India}

\begin{abstract}
 The electroviscous effect deals with the change in the viscosity of fluids due to an external electric field. 
Here, we report experimental studies on the electroviscous effects in a ferroelectric nematic liquid crystal. \SN{It was synthesised accomplishing a new synthetic route which provides higher yield than conventional one}. We measure electric field-dependent viscosity under a steady shear at different temperatures. In the low field range, the increase in viscosity ($\Delta\eta=\eta(E)-\eta_0$) is proportional to $E^2$ and the corresponding viscoelectric coefficient \SD {($f\approx10^{-9}$m\textsuperscript{2}/V\textsuperscript{2}) of the  ferroelectric nematic is 2 orders of magnitude larger than the apolar nematic liquid crystals and largest ever measured for a fluid}. The apparent viscosity measured under a high electric field shows a power-law divergence $\eta\sim(T-T_c)^{-0.7\pm0.05}$, followed by \SD{nearly an order of magnitude drop} below the N-N\textsubscript{F} phase transition.
 Experimental results within the dynamical scaling approximation demonstrate rapid growth of polar domains under a strong electric field as the \SD{N-N\textsubscript{F} phase transition is approached. The gigantic electroviscous effects demonstrated here are important for emerging applications and understanding striking electrohydromechanical effects in ferroelectric nematic liquid crystals.}

 %\SN{The gigantic electroviscous effects demonstrated here are important for understanding striking electrohydromechanical effects and also potential for applications in micro and nanofluidic electromechanical devices (MEMS and NEMS) based on ferroelectric nematic liquid crystals.}}
 
 %\SN{It will be also  useful for micro/nanofluidic electromechanical systems (MEMS/NEMS) based on ferroelectric nematic liquid crystals. }
 %based on striking electrooptics
 %The gigantic electroviscous effects demonstrated here are important for all striking electrooptics and electrohydromechanical effects observed in ferroelectric nematic liquid crystals.
%The gigantic electroviscous effects demonstrated here are important for emerging applications and understanding striking electrohydromechanical effects in ferroelectric nematic liquid crystals.}

 \end{abstract}
\preprint{HEP/123-qed}
\maketitle

%\section{Introduction}
Nematic (N) liquid crystals (LCs) are widely used in flat panel displays. They are apolar i.e, the mean molecular direction, called the director $\hat{n}$ is equivalent to $-\hat{n}$, hence there is no macroscopic polarisation~\cite{pg}. Long ago Max Born predicted the existence of polar nematic i.e, the ferroelectric nematic~\cite{MB}. However, such a phase was elusive for a long time. \SD{Very recently the ferroelectric nematic phase (N\textsubscript{F}) has been discovered in a highly polar with slightly wedge-shaped molecules~\cite{RJM1,RJM2,NH,MA,AM,NS,XC}. It has created a lot of interest in the science and engineering community because of its technological applications and intriguing phenomena important for fundamental science~\cite{MTM1,MTM2,RB,LC,JY,MP,RSa}.} The ferroelectric nematic has high potential to replace the existing non-polar nematics in displays because of its striking electro-optical properties and \SD{fast response time}~\cite{XC,XC1,NS1}. Although, several new physical and electrooptical properties of {N\textsubscript{F} phase have been reported~\cite{SB,OD,HN1,PR,CLF,FC,JLI,XZ} their flow viscosities and the effect of external electric field on the flows are yet unexplored. Since the constituent molecules are highly polar the electroviscous effects in ferroelectric NLCs are expected to be more robust than in ordinary nematic liquid crystals due to the interaction of polarisation with the electric field.
   
The electroviscous effect deals with the change in viscosity especially in polar liquids in the presence of the electric field. The electric field-dependent viscosity of such liquids can be expressed as~\cite{RH,EN2}
\begin{equation}
\eta(E)=\eta_0(1+f|E|^{2})
\end{equation} 
  where $\eta_0$ is the viscosity in the absence of the field and $f$ is the viscoelectric coefficient. For many organic liquids including water, $f\sim10^{-13}-10^{-16}$ m\textsuperscript{2}V\textsuperscript{-2}~\cite{RH,DJ}. When the fluid is a nematic liquid crystal (NLC) the viscosity depends on the orientation of the director with respect to the flow direction. There are three flow viscosity coefficients of nematics which are known as Miesowicz viscosities~\cite{MM,RG}. In the case of $\eta_1$, the director is perpendicular to the flow direction i.e., $\hat{n}\perp \overrightarrow{v}$ and parallel to the velocity gradient i.e., $\hat{n}||\overrightarrow{\nabla} v$. For $\eta_2$, the director is parallel to the flow direction i.e., $\hat{n}||\overrightarrow{v}$. In the case of $\eta_3$ the director is perpendicular to both the velocity and velocity gradient directions, i.e., $\hat{n}\perp\overrightarrow{v}$ and $\hat{n}\perp \overrightarrow{\nabla} v$. Usually for calamitic liquid crystals (rod-like molecules) $\eta_1>\eta_3>\eta_2$~\cite{pg,chim,HI,SC}. These viscosity coefficients depend on the molecular structure as well as on the intermolecular interactions and usually increase with decreasing temperature~\cite{JA,MTC}. 
  %%%%%%%%%%%%%%%
\begin{figure}[!ht]
\begin{center}
%\vskip -0.1in
\includegraphics[scale=0.4]{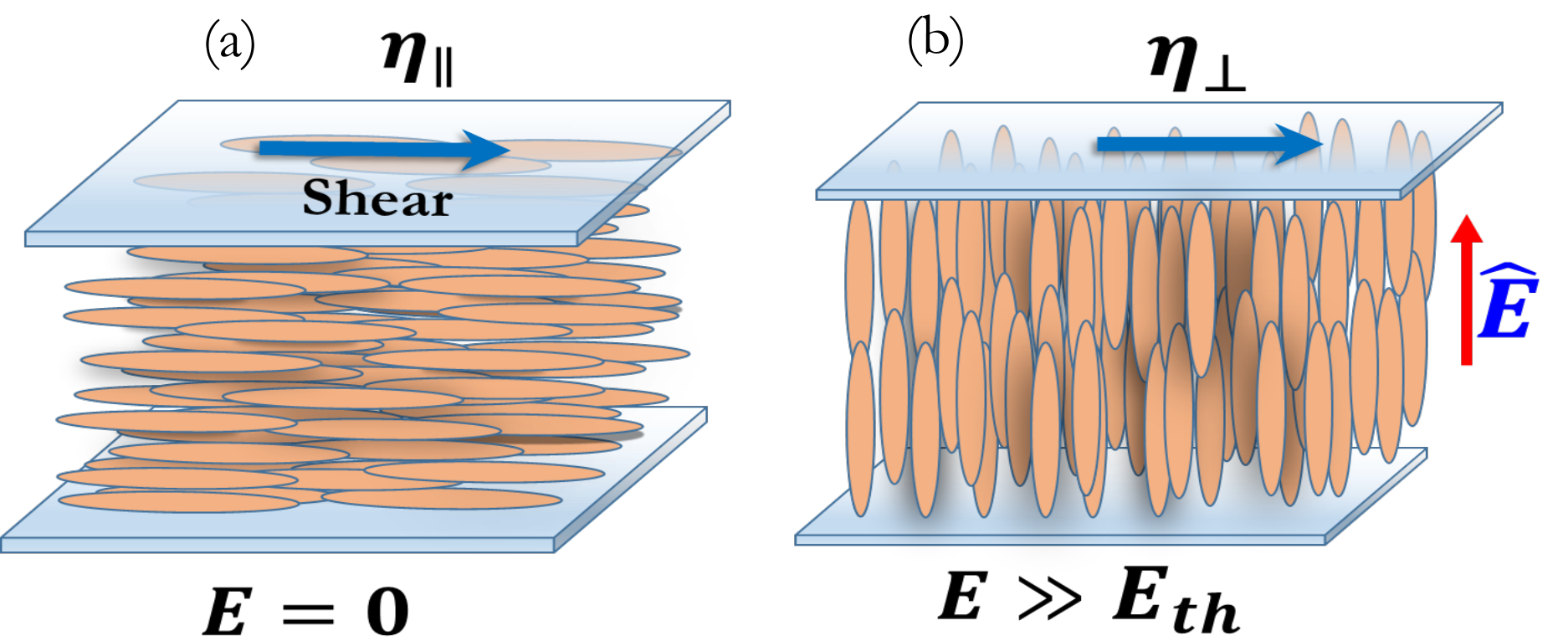}
\caption{ Orientation of the LC director $\hat{\bf{n}}$ without and with electric field, $E$. (a) In the absence of an electric field, $\hat{n}$ is parallel to the \SD{flow direction (blue arrow)} and the apparent viscosity $\eta\simeq\eta_{||}$. (b) Above the Freedericksz threshold field $E>>E_{th}$, $\hat{n}$ is perpendicular to the \SD{flow} direction and the apparent viscosity $\eta\simeq\eta_{\perp}$.
\label{fig:figure1}}
\end{center}
%\vskip -0.2in
\end{figure}
%%%%%%%%%%%%%%

%%%%%%%%%%%%%%%
\begin{figure*}[!ht]
\center\includegraphics[scale=0.6]{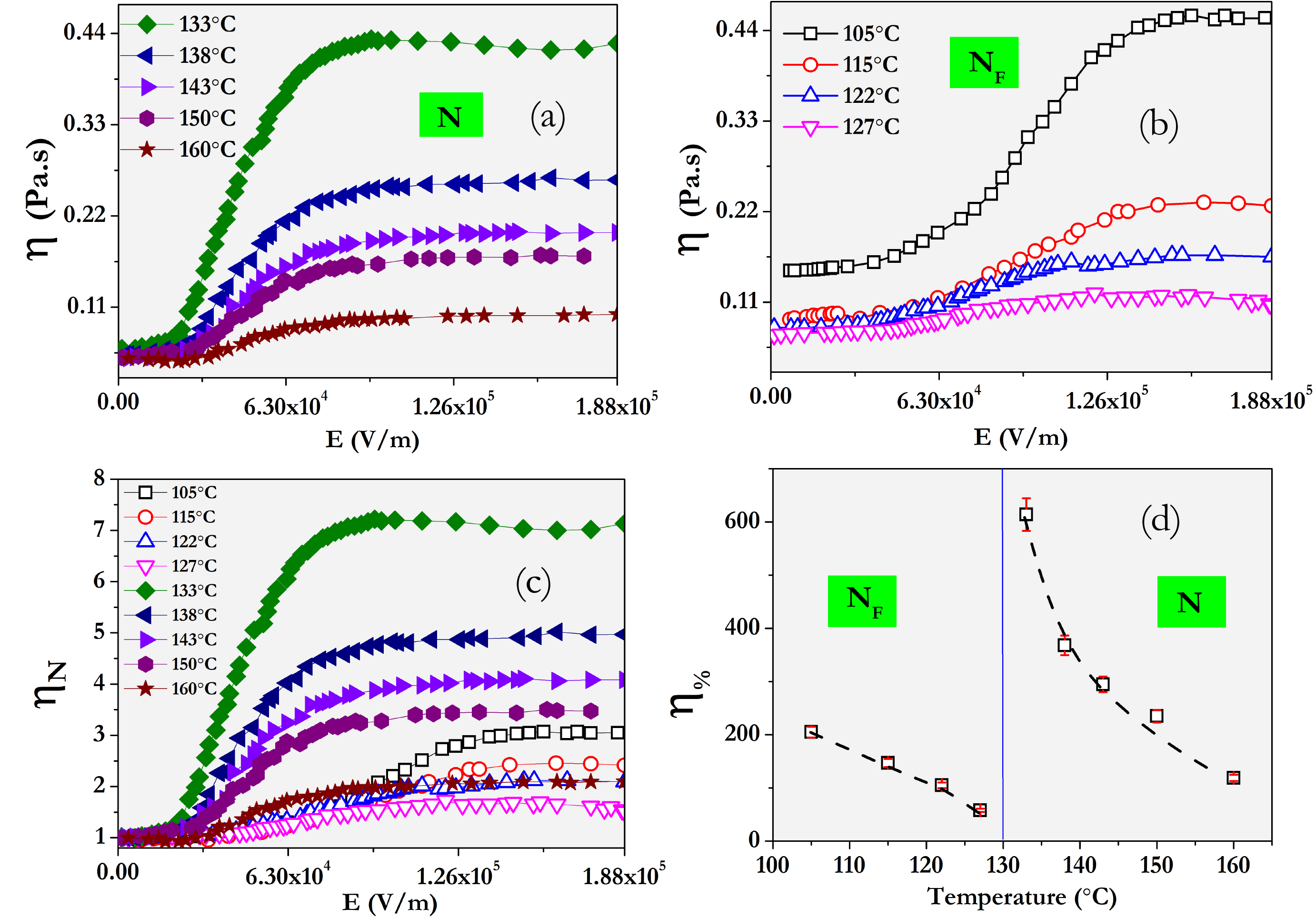}
%\vskip -0.1in
%[height=0.4\textwidth, width=0.9\textwidth]
\caption{ Change of apparent viscosity $\eta$ with field at different temperatures in the \SN{(a) N and (b) N\textsubscript{F} phases}. Measurements are made at a shear rate $\dot{\gamma}=120$ s\textsuperscript{-1}. (c) Normalized viscosity, \SD{$\eta_N=\frac{\eta(E)}{\eta_0}$} in both phases, where $\eta_0$ is the constant viscosity in the absence of the electric field. (d) The percentage increment of viscosity, $\eta_{\%}=\frac{\eta(E=E_s)-\eta_0}{\eta_0}\times 100$, at different temperatures, where $E_s$ is the field above which the viscosity is saturated. We chose $1.7\times10^{5}$ V/m. The vertical line indicates N-N\textsubscript{F} phase transition temperature. Dashed line is a guide to the eye.
\label{fig:figure2} }
%\vskip -0.2in
\end{figure*}
%%%%%%%%%%%%%%%

 There are a few studies on the effect of electric field on the flow viscosities of ordinary NLCs~\cite{PP,JA,MTC,KN,JJ,JA1,JA2}. In the absence of an electric field, the director $\hat{n}$ of the presheared nematic is parallel to the \SD{shear-flow} direction and we can write the apparent viscosity $\eta\simeq\eta_2\simeq\eta_{||}$ (Fig.\ref{fig:figure1}(a)). Preshear is necessary to get a uniform alignment of the liquid crystal director. For a NLC with positive dielectric anisotropy,  above the Freedericksz threshold electric field ($E_{th}$), $\hat{n}$ reorients toward the field direction and the apparent viscosity increases and saturates at higher fields and $\eta\simeq\eta_1\simeq\eta_{\perp}$ (Fig.\ref{fig:figure1}(b)). Here, $\eta_{||}$ and $\eta_{\perp}$ are the viscosities of the NLCs with the director $\hat{n}$ parallel and perpendicular to the flow direction, respectively. The rising part of the viscosity from zero fields to the onset of saturation can be expressed by Eq.(1) and the coefficient $f$ can be regarded as the viscoelectric coefficient of the nematic liquid crystal. It signifies the growth rate of the electroviscosity due to the applied electric field.

 %%%%%%%%%%%%%%%
\begin{figure*}[htbp]
%\vskip -0.2in
\includegraphics[scale=0.6]{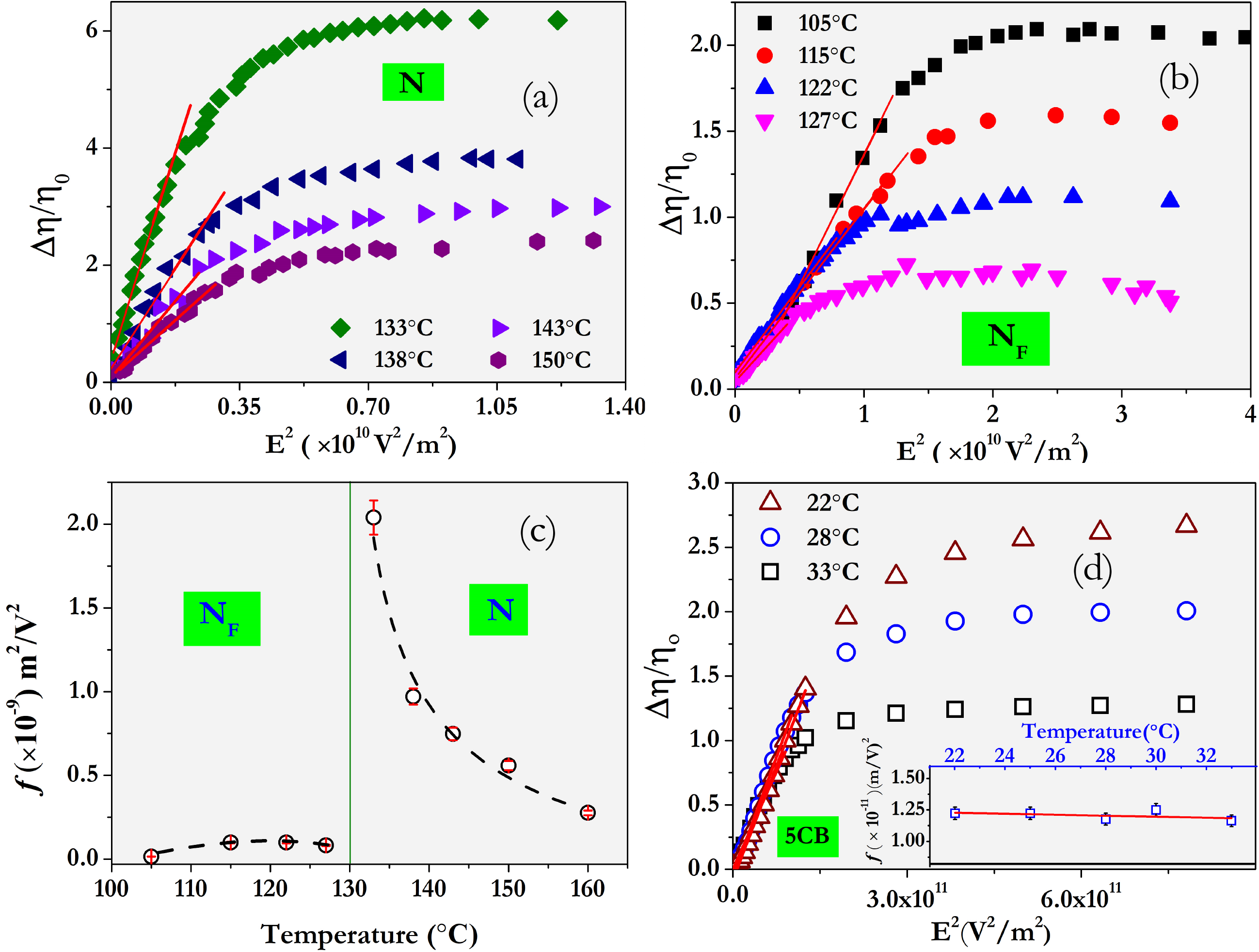}
%[height=0.4\textwidth, width=0.48\textwidth] 
\center\caption{Variation of  $\Delta\eta/\eta_0$ with $E^2$ at different temperatures in (a) N and (b) N\textsubscript{F} phases. The red lines are best fit to the linear part of the curves. (c) Variation of viscoelectric coefficient $f$ with temperature. \SN{Dashed line is a guide to the eye}. The vertical line indicates N-N\textsubscript{F} phase transition temperature. (d) Variation of  $\Delta\eta/\eta_0$ with $E^2$ of 5CB at a few temperatures. Shear rate $\dot{\gamma}=120$ s\textsuperscript{-1}. Inset shows variation of $f$ with temperature of 5CB LC.
\label{fig:figure3}}
%\vskip -0.1in
\end{figure*}
%%%%%%%%%%%%%%%

 %%%%%%%%%%%%%%%
\begin{figure*}[!htp]
\center\includegraphics[scale=0.6]{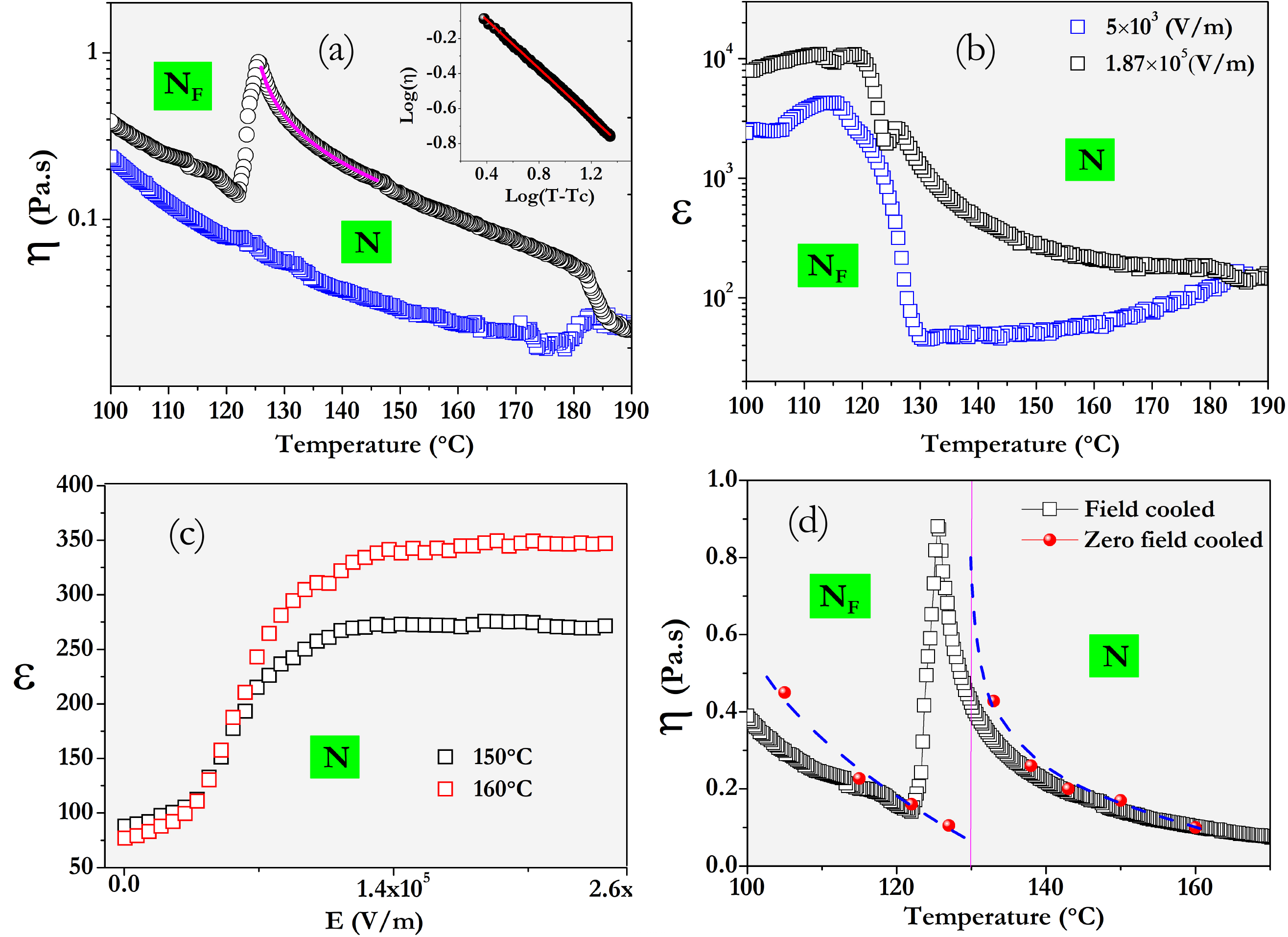}
%[height=0.4\textwidth, width=0.5\textwidth] 
\caption{ (a) Temperature dependence of apparent viscosity $\eta$ measured at $\dot{\gamma}=120$ s\textsuperscript{-1}, at two electric fields.  Blue squares correspond to data measured at $E=5\times10^{3}$ V/m ($<<E_{th}$) and  black circles correspond to data measured at  $E=1.87\times10^{5}$ V/m ($>> E_{th}$).  Inset shows a power-law fit to $\eta\sim (T-T_C)^{-0.7\pm0.05}$ to the diverging viscosity where $T_c=124.13^\circ$C. \SD{ Magenta curve indicates the temperature range selected for fitting.}
 (b) \SD{Temperature dependence of effective dielectric constant} $\epsilon$  measured \SN{at $f=$ 600 Hz}. (c) \SD{Variation of $\epsilon$ with field} at two temperatures in the N phase. (d) Growth of viscosity $\eta$ as a function of temperature in two different conditions. One with the electric field $E$ on. This is referred to here as ``field-cooled'' (black squares, $E=1.87\times10^{5}$ V/m) and the other set (red dots) are obtained from Fig.\ref{fig:figure2}(a,b) where the system is cooled to the lower temperature first and then the field is applied to measure the viscosity. Dashed blue line is a guide to the eye for zero-field cooled data. The vertical line indicates  N-N\textsubscript{F} transition temperature under zero-field. 
 %The vertical line indicates the N-N\textsubscript{F} phase transition temperature $130^\circ$C in the absence of an electric field.
  \label{fig:figure4}}
 %  \vskip -0.1in
\end{figure*}
%%%%%%%%%%%%%%

\section{Results and Discussion}
%\section{Results and discussion}
We work with a ferroelectric nematic liquid crystal RM-734, first reported by Mandle \textit{et al.}~\cite{RJM1}. We have developed a new method for diester liquid crystal synthesis. It is based on the inversion of molecular core expansion (see Materials section). With such an approach, the overall yield of the synthesis is increased and the possibility of obtaining highly fluorinated liquid crystal diesters, especially those bearing fluorine atoms in the central phenyl ring is enhanced. The description of our novel approach and the synthesis summary are presented in the supplementary~\cite{sup}. 
\SD{We set up a computer-controlled electrorheological experiment based on a rheometer. The measuring device consists of two parallel plates with a fixed gap connected to an LCR meter and interfaced with a computer. The setup is capable of measuring the viscosity and the dielectric constant simultaneously (supplementary information~\cite{sup}).}
%\SD{It} allows us to directly measure the \SD{viscoelectric} coefficient.
 First, we measured electric field-dependent viscosity, known as apparent viscosity at different temperatures \SD{by changing the amplitude of the electric field at a fixed frequency}. The viscosity of the presheared sample is independent of the shear rate (SI)~\cite{sup}, hence we fixed the shear rate $\dot{\gamma}=120$s\textsuperscript{-1} for all measurements. Figure \ref{fig:figure2}(a) shows the variation of apparent viscosity $\eta$ at different temperatures in the nematic (N) phase of RM-734. The viscosity below a particular field (Freedericksz threshold, $E_{th}\simeq2.5\times10^4$ V/m) is constant ($=\eta_0$) and it rises rapidly with the field and saturates above $E\simeq 1.0\times10^{5}$ V/m. Figure \ref{fig:figure2}(b) shows the variation of apparent viscosity at different temperatures in the N\textsubscript{F} phase. In this phase, the viscosity shows similar trends but the relative enhancement of viscosity with respect to the zero field viscosity is smaller compared to that of the N phase. In order to bring out the difference we introduce normalised viscosity, defined as \SD{$\eta_N=\frac{\eta(E)}{\eta_0}$}, where $\eta_0$ is the zero-field viscosity (Fig.\ref{fig:figure2}(c)). It is noted that the change of saturated viscosity with temperature is non-monotonic. For example, the saturated viscosity increases as the temperature is raised from 160$^\circ$C and reach a maximum value at the N-N\textsubscript{F} transition and then decreases drastically as the temperature is reduced. 
  Figure \ref{fig:figure2}(d) shows the relative enhancement of viscosity in terms of percentage, which can be written as $\eta_{\%}=\frac{\eta(E=E_s)-\eta_0}{\eta_0}\times 100$, where $\eta(E=E_s)$ is the saturated viscosity. The percentage increment of viscosity $\eta_{\%}$ increases rapidly as the N-N\textsubscript{F} phase transition is approached and the pronounced effect is observed very close to the N-N\textsubscript{F} transition temperature where the enhancement is about 600\%. Below the phase transition, it drastically decreases to 50$\%$ (T= 127$^{\circ}$C) and again gradually increases to 200$\%$ which will be discussed later.

   We have obtained the viscoelectric coefficient $f$ at different temperatures using Eq.(1) which can be expressed as $\Delta\eta/\eta_0=fE^2$, where $\Delta\eta=\eta(E)-\eta_0$. Figure \ref{fig:figure3}(a) and  (b) shows the variation of $\Delta\eta/\eta_0$ with $E^2$ in N and N\textsubscript{F} phases, respectively. The coefficient $f$ is obtained from the slope of the linear part of the curve in the low field region (below the onset of saturation) and shown in Fig. \ref{fig:figure3} (c). In the N phase (e.g., T=160$^{\circ}$C) $f=0.4\times10^{-9}$ m\textsuperscript{2}V\textsuperscript{-2} and it increases rapidly as the N-N\textsubscript{F} transition is approached. The maximum value is obtained ($f=2.1\times10^{-9}$ m\textsuperscript{2}V\textsuperscript{-2}) near N-N\textsubscript{F} transition temperature. This is about six orders of magnitude larger than that of ordinary liquids like water~\cite{RH,DJ} \SD{and the largest ever measured for a fluid.}

    In order to compare the same with ordinary NLCs with polar molecules we measured the viscoelectric coefficient of 5CB (pentyl cyanobiphenyl) liquid crystal. Figure \ref{fig:figure3}(d)) shows a change in apparent viscosity with the electric field at a few temperatures and corresponding $f$ of 5CB LC. The coefficient $f$ ($\simeq1.25\times10^{-11}$ m\textsuperscript{2}V\textsuperscript{-2}) is  independent of temperature and almost two orders of magnitude smaller than  RM-734 liquid crystal. The molecules of 5CB have an axial dipole moment of 4.8 D~\cite{MC} which is almost two and a half times smaller than compound RM-734 (11.3 D)~\cite{MA}. Thus, due to the large dipole moment and resulting dipole-dipole correlation, the viscoelectric coefficient of RM-734 is expected to be much higher than nematics with nonpolar or weakly polar molecules. However, two orders of magnitude larger $f$ of RM-734 and its rapid growth indicate the emergence of polar domains that grows rapidly as the N-N\textsubscript{F} transition is approached.

 In what follows, we have measured the temperature-dependent apparent viscosity $\eta$ of a  presheared sample.  We fixed the shear rate $\dot{\gamma}=120$ s\textsuperscript{-1}, and measured the viscosity $\eta_{||}$ ($E<<E_{th}$) and $\eta_{\perp}$ ($E>>E_{th}$) as a function of temperature (Fig.\ref{fig:figure4}(a)). The viscosity increases with decreasing temperature as expected. No significant change in $\eta_{||}$ is observed at the N-N\textsubscript{F} transition temperature ($T=130^{\circ}$C). On the other hand $\eta_{\perp}$ increases much more rapidly with decreasing temperature and tends to diverge at the N-N\textsubscript{F} transition temperature. \SD{It reaches a maximum of 900 mPas at the transition and the corresponding viscosity anisotropy $\delta\eta=\eta_{\perp}-\eta_{||}=750$ mPas,  which is nearly an order of magnitude larger than the ordinary NLCs~\cite{chim,KN,JA} and largest viscosity anisotropy ever measured for any NLC.} It is noted that $\eta_{\perp}$ changes slope \SD{below} temperature T=150$^\circ$C, at which the collective behaviour of the molecules starts to develop~\cite{AM}.  Below the N-N\textsubscript{F} transition the viscosity drops down significantly to about 140 mPas from 900 mPas, showing $\eta_{\perp}$ of the N\textsubscript{F} phase is reduced compared to the N phase although the orientational order parameter of the N\textsubscript{F} phase is larger~\cite{RJM2}. 
 \SD{Such a drastic decrease of $\eta_{\perp}$ is consistent with the first-order nature of the N-N\textsubscript{F} phase transition and it suggests that microscopically the structure of these two phases are different.} 
   \SD{ Although the dielectric properties of this LC are quite complex~\cite{D1,D2,D3} we could simultaneously measure an effective dielectric constant $\epsilon$ (Fig.\ref{fig:figure4}(b)) at a frequency of \SN{600 Hz}~\cite{note2}.}  Since the dielectric anisotropy is positive, $\epsilon\simeq\epsilon_{\perp}$ when $E<<E_{th}$ and $\epsilon\simeq\epsilon_{||}$ when $E>>E_{th}$ (Fig.\ref{fig:figure4}(c)).   It is noted that $\epsilon_{||}$ shows a nearly diverging trend followed by a small kink at the N-N\textsubscript{F} phase transition temperature (Fig.\ref{fig:figure4}(a)). The diverging trend in $\epsilon_{||}$ is a signature of cooperative molecular motion possibly leading to polar domains \SD{which are basically dynamic clusters of parallel dipoles.} %\SD{that contributes to the viscosity}. 

 %%%%%%%%%%%%%%%
%\begin{figure}[!htp]
%\center\includegraphics[scale=0.55]{Fig5.pdf}
%\caption{ Growth of viscosity $\eta$ as a function of temperature, in two different conditions. One with the electric field $E$ on. This is referred to here as ``field cooled'' (black squares, $E=1.87\times10^{5}$ V/m) and the other set (red dots) are obtained from Fig.\ref{fig:figure2}(a,b) where the system is cooled to the lower temperature first and then the field is applied to measure the viscosity. The dashed blue line is a guide to the eye for zero-field cooled data. The vertical line indicates  N-N\textsubscript{F} transition temperature under zero-field. 
%The transition temperature under field cooled conditions is reduced by $4.5^\circ$C as evident from the viscosity peak.
 % \label{fig:figure5}}
 %  \vskip -0.2in
%\end{figure}
%%%%%%%%%%%%%%

  At high temperatures, due to thermal fluctuations, the size of the polar domains should be small, but as the temperature is decreased towards N-N\textsubscript{F} transition, the domain size will grow. \SD{Higher electric field can enhance parallel correlation of dipoles and facilitate domain growth.  Under these conditions, the viscosity is expected to increase, as seen in Fig.\ref{fig:figure4}(a) because larger stress will be needed to deform the polar domains.} Thus, one does not expect a strong deviation between the measurements of viscosity done in the following two scenarios; (1) with the high field on at high temperature and then cooled to the lower temperature (referred to here as field cooled) and (2) first cooled at the required temperature and then apply the desired electric field (zero field-cooled) and measure the viscosity. In both these scenarios, the domain will grow when the electric field is applied in the N phase, so one expects the measurements to be very similar qualitatively and quantitatively. As shown in Fig.\ref{fig:figure4}(d), the data indeed suggest the same (see data for $T>T_c$).
The results will be very different below the transition temperature simply because at zero field cooling conditions; the system will go through a sharp structural change due to N to N\textsubscript{F} phase transition. On the other hand, in the field-cooled conditions, the polar domains in the N phase will not allow the system to immediately go to the N\textsubscript{F} phase and the sharp phase transition will \SD{be rounded as evident from Fig.\ref{fig:figure4}(d).} 
Although the scenario proposed seems to be quite likely the reason for the observation, we do not have direct proof of the growth of the polar domain with the applied field. Future computational studies on these systems can unearth the microscopic mechanisms for these experimental observations.

\SK{If we compare the temperature-dependent viscosity (Fig.\ref{fig:figure4}(a)) and dielectric constant  (Fig.\ref{fig:figure4}(b)) measured at the high field and those at the low field  then N-N\textsubscript{F} transition temperature at high field is found to be decreased by about $3.5^\circ$C. \SD{This decrease is not due to the sample degradation under field~\cite{x}.} It can be rationalized as caused by the structural changes during the N-N\textsubscript{F} phase transition following two possible scenarios. Firstly, the high field  can suppress the pre-transitional splay fluctuations~\cite{MA,tobe} and favours domain growth with the polarity along the field and hence one would expect the N-N\textsubscript{F} transition temperature to decrease if it is considered that the ferronematic phase can have domains with splay structure. Note that the largest applied field in this experiment is not sufficient to align all the molecules along the field direction in the N\textsubscript{F}, otherwise, there will not be any phase transition. Secondly, the polar regions will grow in size but they can not span the entire sample as the nematic phase does not support the orientation of all the molecules in the same direction due to large dipole moments. With decreasing temperature, these polar regions grow in size and thus we see a power law change in the measured viscosity in the nematic phase in field-cooled condition. As the temperature is lowered below the equilibrium N-N\textsubscript{F} transition, thermodynamically the N\textsubscript{F} phase with opposite polarity domains will be favoured as observed in the experiment \cite{XC}, but the already formed polar regions will oppose the formation of this N\textsubscript{F} phase until at a lower temperature where the enthalpy will win over and the system will go to the N\textsubscript{F} phase.  The near equality of the viscosity  data measured in the field-cooled and zero-field-cooled scenarios below the N-N\textsubscript{F} phase transition indeed supports this argument.  Note that \SD{in the latter scenario the}   splay structure is not essential to explain the observed decrease in the transition temperature. }
%\PK{However, further experimental and theoretical studies 
%are required to have a better understanding of the microscopic mechanism.} }
 
  %%%%%%%%%%%%%%%
\begin{figure*}[htbp]
%\vskip -0.2in
\includegraphics[scale=1]{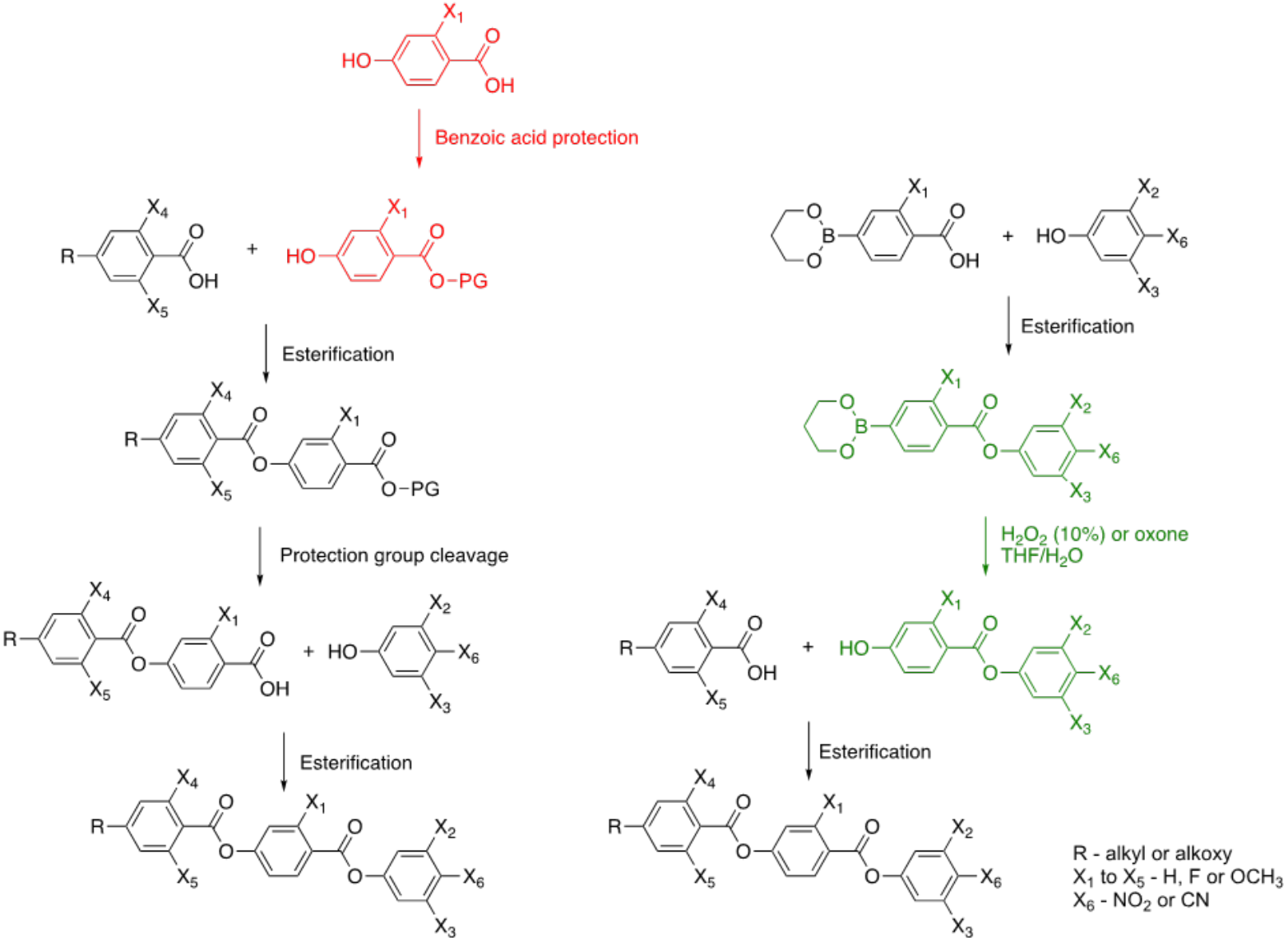}
%[height=0.4\textwidth, width=0.48\textwidth] 
\center\caption{ Comparison of novel synthetic methodology for RM-734 and its fluorinated analogues (right synthetic route) with conventional one (left synthetic route).
\label{fig:figure5}}
%\vskip -0.1in
\end{figure*}
%%%%%%%%%%%%%%%

   Further, it is expected that there will be a growth of a length scale related to the formation of polar domains in the nematic phase and if we assume that the length scale shows a divergence at the transition temperature, $T_c$ as $\xi\sim |T-T_c|^{-\nu}$ with $\nu$ being the correlation length exponent, then one can expect to see a similar power-law divergence of the viscosity within the dynamical scaling approximation \cite{PCH} as $\eta\sim \xi^{z}\equiv |T-T_c|^{-\nu z}$ where $z$ is the dynamical scaling exponent. The power-law fit to the viscosity date is shown in the inset of Fig.\ref{fig:figure4} (a), and the fit is very nice to confirm such a possible power-law divergence of viscosity with an exponent $\nu z\sim0.7$. Currently, we do not have an independent estimate of the exponent $\nu$, so we can not compute the exponent $z$. On the other hand, the correlation length exponent in critical phenomena within mean field approximation is $\nu=\frac{1}{2}$, so if we assume the same exponent here, then the dynamical exponent $z\sim1.4$. This suggests that the domains are probably more elongated (one-dimensional) in nature than spherical. Some of these conjectures can be tested via computer simulation studies which are in progress.
  
 \SD{ We would like to make a comment here. A one-dimensionally modulated splay structure of the ferroelectric nematic phase of compound RM-734 was proposed by Mertelj \textit{et al.}~\cite{MA} whereas a two-dimensionally modulated splay structure is proposed by the theory~\cite{MP}. However, Chen \textit{et al.} reported the occurrence of spontaneously polar domains of the opposite sign of polarisation separated by distinct domain boundaries~\cite{XC}. Our explanation on the temperature-dependent electroviscosity does not rely on any particular structure of the ferroelectric nematic phase. }

%\section{Conclusion} 
To summarise, ferroelectric nematic LCs exhibit giant \SD{electroviscous effects and the viscoelectric coefficient obtained is the largest ever measured for a fluid}. The apparent viscosity under field-cooled \SD{condition shows a power-law divergence as the N-N\textsubscript{F} transition is approached followed by a drastic decrease below the transition.} 
%\SD{A significant viscosity drop below the N-N\textsubscript{F} phase transition temperature under a high electric field suggests that microscopically the structures of the respective phases are different. }
%We hope our results will be useful in establishing the true structure of ferroelectric nematic LC which is still debated~\cite{AM,XC,MP,RSa}.
 Large viscoelectric coefficient and pretransitional divergence of the electroviscosity indicate a strong polar correlation resulting in elongated domains that grows rapidly with decreasing temperature. 
 %The effect of such collective molecular response is also evident in the temperature-dependent effective dielectric constant.
  We envisage that the pretransitional growth of polar domains and their dynamic response should be manifested in many other physical properties and effects. \SN{The new synthetic route reported here is significant for the large-scale production which is crucial from the application and broad study perspectives.  These new results obtained are important for most of the physical, electrooptical as well as the electrohydromechanical~\cite{MTM1,RB,MTM2,LC} effects in ferroelectric nematic liquid crystals.}
 \SN{They may also be useful for applications in micro and nanofluidic electromechanical devices (MEMS/NEMS) based on ferroelectric nematic liquid crystals. }\\

\section{Materials}
We have developed a new approach for the synthesis of nitro and cyano diesters known for exhibiting ferroelectric character. We have used the opposite direction of molecular core expansion, compared to the known synthetic method~\cite{RJM1}. In this method, successive molecular parts are introduced to the molecule concerning the direction of its dipole moment. Four main steps describing this molecular approach are benzoic acid protection, esterification with the appropriate benzoic acid, carboxylic protective group cleavage and, finally, the last esterification reaction. The key half-product for obtaining the diester derivatives is the eligibly substituted 4-hydroxybenzoic acid. Thus, the protection of carboxylic acid is essential. The benzyl group is used widely for this purpose. This group is not fully selective towards the carboxylic acid, leading to the mixture of protected carboxyl and hydroxyl groups, especially in the case of fluorinated analogues. The yield of the protection of unsubstituted 4-hydroxybenzoic acid is approximately 50\%~\cite{DZ}. By introducing the fluorine atom to 4-hydroxybenzoic acid, the yield of the protection reaction significantly decreases~\cite{SB}. It is because of increasing the acidity of the hydroxyl group caused by the inductive effect. The protection reaction is no longer selective. It limits the use of this method in the case of highly fluorine substituted compounds (Fig.\ref{fig:figure5}).
In our method, the molecular core expansion direction is inverted. Instead of using a half-product without the regioselectivity towards one of the functional groups, we have used one with a ``hidden'' precursor of the hydroxyl group. Three main steps characterise our method: esterification of key half-product with an appropriate nitro or cyano phenol, introduction of the hydroxyl group and the final esterification reaction. The key half-product bears a boronic ester group as a ``hidden'' precursor of the hydroxyl group, which can be performed by the boronic ester oxidation in mild conditions~\cite{book}. This reaction, carried out in mild conditions, leads to the corresponding phenol with a high yield. The details of the synthetic scheme of RM-734 is presented in the supplementary~\cite{sup}. \\\\

\textbf{Acknowledgments}: SD acknowledges financial support from SERB-SUPRA (Ref. No: SPR/2022/000001). MPK acknowledges UGC-CSIR for fellowship. This work is co-financed by the European Social Fund under the``Operational Programme Knowledge Education Development 2014-2020''. We acknowledge N. V. Madhusudana from Raman Research Institute  for the useful discussions. SK acknowledges funding by TIFR (Ref. RTI 4007) and DST (Ref. DST/SJF/PSA-01/2018-19) and SB/SFJ/2019-20/05. PK acknowledges UGB 22-801 project.

\end{document}